\documentstyle[12pt,epsf,epsfig]{article}
\def\be{\begin{equation}}
\def\ee{\end{equation}}
\def\ba{\begin{eqnarray}}
\def\ea{\end{eqnarray}}

\def\bdm{\begin{displaymath}}
\def\edm{\end{displaymath}}
\def\la{~\mbox{\raisebox{-.6ex}{$\stackrel{<}{\sim}$}}~}
\def\ga{~\mbox{\raisebox{-.6ex}{$\stackrel{>}{\sim}$}}~}
\def\bq{\begin{quote}}
\def\eq{\end{quote}}

 at 10truept

\newcommand{\beq}{\begin{equation}}
\newcommand{\eeq}{\end{equation}}
\newcommand{\bea}{\begin{eqnarray}}
\newcommand{\eea}{\end{eqnarray}}
\newcommand{\beqa}{\begin{eqnarray}}
\newcommand{\eeqa}{\end{eqnarray}}

\def\la{~\mbox{\raisebox{-.6ex}{$\stackrel{<}{\sim}$}}~}
\def\ga{~\mbox{\raisebox{-.6ex}{$\stackrel{>}{\sim}$}}~}

\def\ltap{\ \raise.3ex\hbox{$<$\kern-.75em\lower1ex\hbox{$\sim$}}\ }
\def\gtap{\ \raise.3ex\hbox{$>$\kern-.75em\lower1ex\hbox{$\sim$}}\ }
\def\gl{\ \raise.5ex\hbox{$>$}\kern-.8em\lower.5ex\hbox{$<$}\ }
\def\roughly#1{\raise.3ex\hbox{$#1$\kern-.75em\lower1ex\hbox{$\sim$}}}

\begin{document}

\thispagestyle{empty}
\begin{flushright}
arXiv:0904.2394 [astro-ph.CO]\\
April  2009
\end{flushright}
\vspace*{1.75cm}
\begin{center}
{\Large \bf Levitating Dark Matter}\\

\vspace*{1.2cm} {\large Nemanja Kaloper$^{a,}$\footnote{\tt
kaloper@physics.ucdavis.edu} and Antonio Padilla$^{b,}$\footnote{\tt
antonio.padilla@nottingham.ac.uk} }\\
\vspace{.5cm} {\em $^a$Department of Physics, University of
California, Davis, CA 95616, USA}\\
\vspace{.2cm} {\em $^{b}$School of Physics and Astronomy, 
University of Nottingham, Nottingham NG7 2RD, UK}\\

\vspace{1.25cm} ABSTRACT
\end{center}
A sizable fraction of the total energy density of the universe may be in heavy particles with a 
net dark $U(1)'$ charge comparable to its mass. When the charges 
have the same sign the cancellation between their gravitational and gauge forces may lead to a 
mismatch between different measures of masses in the universe. Measuring galactic masses by orbits 
of normal matter, such as galaxy rotation curves or lensing, will give the total mass, while the flows of 
dark matter agglomerates may yield smaller values if the gauge repulsion is not accounted for. 
If distant galaxies which house light beacons like SNe Ia contain such dark particles, the 
observations of their cosmic recession may mistake the weaker forces for an extra `antigravity', 
and infer an effective dark energy equation of state smaller than the real one. In some cases, including that of a cosmological constant, these effects can  mimic 
$w<-1$. They can also lead to a {\it local} variation of galaxy-galaxy forces, yielding a
larger `Hubble Flow' in those regions of space that  could be taken for a dynamical 
dark energy, or superhorizon effects. 

\vfill \setcounter{page}{0} \setcounter{footnote}{0}
\newpage

\section{Introduction}

A variety of cosmological observations has revealed the prevalence of a dark sector in our universe. Typically we model it as a dark energy component, comprising some $73\%$ of the total energy contents, and a dark matter component, contributing another $23\%$ of the energy, with the rest in the form of familiar baryons (see, e.g., \cite{wmap}). The simplest models of dark energy and dark matter posit that they are just two separate entities, most often a cosmological constant and a WIMP particle, in stark contrast to the visible sector which is known to contain a plethora of stable modes. This may at least make us curious that whatever birthed our universe may have been just as creative in populating the dark sector, as it has been in the visible sector. 

Many of our theories would appear to suggest this possibility. The additional dark modes are not yet 
necessitated by direct observational evidence\footnote{See, however, \cite{farros} for an interesting suggestion.}. However, promising micro-physical theories such as those which arise in string theory, for example, admit a dark population of the universe which may be at least as diverse as the visible one. There are many candidates for dark matter particles with a variety of interactions between them, with large masses and weak direct couplings to the visible matter. Among them some may indeed behave just like the WIMP, and be a dominant component of the relic dark matter abundance. In this work we shall consider what happens if it is not alone, focusing on signatures of
additional dark matter with long range interactions, which for simplicity we model by a Maxwell-like theory. Various aspects of such theories have been studied  in the literature \cite{okun}-\cite{ack}, where dark matter was taken to carry either visible or dark charges, and where `dark electrodynamics' have been dubbed $U(1)'$ theories. They are not excluded as long as the dark charges are 
small enough to avoid too much cooling of virialized dark matter in galactic halos, which prevents their collapse into the galactic plane and preserves the observed halo structure. For TeV-mass particles, the bound on the dark fine structure constant is $\alpha' \la 10^{-3}$, with stronger couplings for heavier particles, allowing for quite a large parameter space for such theories \cite{ack}.

Here we are interested in extremely weak dark matter couplings. From the phenomenological point of view, clearly there is no limit as to how weak these couplings might be. However, theoretical arguments based on the considerations of UV completions of theories with weak gauge forces, namely holographic considerations and charged black hole (in)stability, point to the conclusion that in a given effective field theory regime, gauge forces should not be weaker than gravity \cite{nima}. These arguments also suggest that if the gauge theory coupling strength is fixed at $g$, its effective description should break at a low energy cutoff 
$\Lambda \la g M_{Pl}$. Furthermore, there ought to be  stable particles lighter than $\Lambda$, which obey the inequality $M \la Q M_{Pl}$, and may saturate it if they arise from some fundamentally supersymmetric theory.  

In what follows, we will take the $U(1)'$ couplings to be of gravitational strength, as is frequently the case in string theory. This yields $g \sim 10^{-15}$, and so by the arguments of \cite{nima}, it implies 
the existence of a $U(1)'$-charged lightest stable particle in the dark sector, with 
mass $m \la \Lambda \sim 10^{-15} M_{Pl} \sim {\rm TeV}$.
If this particle also has short range weak scale interactions with the Standard Model and/or the remaining dark sector,  by the arguments of \cite{feng}, it may end up with the right relic abundance to contribute to the dark matter in our universe.

If such particles inhabit our universe, they would have a number of interesting cosmological consequences. 
To leading order, some of them would be similar to those already explored in \cite{josh,Gubser,marc}, where an additional Yukawa force due to a scalar exchange was assumed to operate in the dark sector.
The ensuing bounds on the coupling strength, originally quoted to be of the order of gravitational coupling \cite{josh}, were recently pushed down to about 10\% of gravitational strength \cite{marc},  with the underlying assumption that all of the dark matter experiences this extra force. Clearly, the bounds will be correspondingly weaker if only {\it a fraction} of the dark matter carries dark $U(1)'$ charges. In this work we will explore what happens if all the $U(1)'$ charges have the same sign\footnote{This would clearly require some early universe dynamics to explain a net non-vanishing charge in our borough of the universe, such as a $U(1)'$ charged inflaton condensate.}. We should note that the consequences of a weak unbroken $U(1)'$ in a neutral universe have been examined recently in \cite{ack}, where the authors considered interesting questions of dark chemistry and plasma dynamics\footnote{It is, of course, rather difficult to imagine that such issues will be very important in our case when the binding force is as weak as gravity, since the binding energies in the `dark sector' in our case will be $g^4 \sim 10^{-60}$ times smaller than those in the normal sector, so that the bound states just won't be very bound.}. In the case we consider, with feeble $U(1)'$ of dominant strength, the effects one expects should be primarily competitive with gravitational phenomena.  The cancellation between their gravitational and gauge forces may yield a mismatch between the masses 
measured by $U(1)'$ singlets, such as baryon aggregates, and $U(1)'$-charged objects, such as distant galaxies that may contain charged dark matter. Indeed, normal baryonic matter will follow orbits that are specified by the gravitational field alone, enabling us to correctly determine the total mass of the source, as is the case when studying galactic rotation curves or lensing. In contrast, the motion of charged dark matter agglomerates is controlled by both the gravitational field and the dark gauge repulsion.  If one does not account explicitly for the gauge repulsion in the orbit determination, one would underestimate the mass of the charged source. 

This phenomenon directly affects the toils of an observer who extracts the Hubble rate from the cosmic recession of distant galaxies, where the standard candle SNe Ia reside alongside charged dark matter particles.  If the observations are calibrated to the same value of $H_0$ obtained from, e.g., CMB or other visible matter observations, our observer may mistake the weaker net forces for an extra `antigravity'. This could lead to an interpretation involving effective dark energy equations of state smaller than the real one, including mimicking $w<-1$ without any ghosts, phantoms or other similar pathologies. It could also 
cause a mismatch between the measurements of the dark matter density fraction $\Omega_{DM}$ using 
purely baryonic probes of dark matter and probes that track the evolution of total dark matter distributions, which may contain charged dark matter particles. The former are, of course, to leading order blind to dark matter charges,  whereas the latter are  sensitive to dark gauge repulsion. Further, galaxy-galaxy forces may change locally if the density of charged dark matter varies from place to place. This will affect recession velocities, causing deviations from the mean `Hubble value' that could be taken for a dynamical dark energy. Such dynamics might be a useful model to explain recent claims of the observed 
bulk flows in \cite{dale}.

\section{Faradayan cosmology}

To demonstrate our claims, we start with a simple, yet sufficiently illustrative argument involving  a derivation of the Hubble equation using Newtonian cosmology with charges. In other words, a {\it Faradayan} cosmology! This is in fact precisely the same approach that 
Lyttleton and Bondi used in 1959 to argue that a small fractional charge difference between an electron and a proton can yield Hubble expansion in the Steady State universe \cite{bondi}. Indeed, if the charges of the electron and the proton had not cancelled each other exactly, hydrogen atoms would have had a small monopole charge $q$. Let us consider what happens if we take the cosmic matter to be charged under some $U(1)$ field, with all charges having the same sign.
Assuming for the moment that they fill the universe uniformly\footnote{Here we ignore the paradoxes one faces in dealing with uniform distributions of charge in an infinite static space, with wisdom after the fact that such issues may be naturally resolved in time-dependent spacetimes such as an expanding universe, where we can treat a uniform distribution of `charges' (a.k.a masses) in a perfectly meaningful way as an expanding universe.}, we model the extragalactic recession by the Hubble law
$\vec v = H \vec R$,  where $R$ is the distance between the observer and a receding beacon. Describing the effects of gravitational and gauge forces by Gauss's theorem, such that the contribution to net acceleration by sources outside of the sphere centered at the observer's location vanish, the net contributions to acceleration comes from interior sources, and are given by Newton's and Coulomb's law, 
$- G_N\frac{mM}{R}$ and $\frac{qQ}{R}$, respectively.  We can now use total energy conservation to relate the Hubble parameter $H$ to the matter contents inside the sphere, which is known to reproduce the correct General-Relativistic Friedmann equation. Taking the distant beacon to be marginally bound, with total energy $E=0$ corresponding to a spatially flat universe \cite{bondi,bondih,alvarez,jordan}, a little straightforward algebra yields
\be
H^2 = \frac{8\pi G_N}{3} \, \rho_{\rm total} - \frac{8 \pi}{3} \Bigl(\frac{q}{m}\Bigr)_{\rm probe} \,
\Bigl(\frac{q}{m}\Bigr)_{\rm background} \,\,\, \rho_{\rm charged} \, ,
\label{friedmann} \ee
where $\rho_{\rm total}$ is the total cosmological mass density, $\rho_{\rm charged}$ is the mass density of charged particles, $ \Bigl(\frac{q}{m}\Bigr)_{\rm probe}$ is the charge-to-mass ratios of the distant beacon whose recession velocity is probing the cosmic contents, and $\Bigl(\frac{q}{m}\Bigr)_{\rm background}$ is the mean charge-to-mass ratio of the charged fraction of cosmic mass. 

Note that the correct interpretation of this equation is as follows: an observer would measure $H$, extracting it from the recession velocity of a distant object. If the object were neutral, one would simply get the true $\rho_{total}$ of the universe. If the object were charged, on the other hand, and the observer did not account for that effect explicitly, one would need to {\it increase} $\rho_{\rm total}$,  in the regime of distances probed by the observation, relative to what would have been seen by a neutral probe in order to get the same value $H$. Said in yet another way, recall that the Hubble velocity measures the velocity of a particle that is marginally bound to a system of masses, at a location $\vec y$ from the observer who is pinned to a fixed location in the mass fluid. This means that if a probe particle is neutral it must have {\it greater} velocity in order to escape than a charged particle at the same distance. Essentially, the Hubble velocity is the analogue of the Newtonian escape velocity, and it will be smaller if some of the attracting masses are screened by the gauge charges.

Now, clearly, if $\Bigl(\frac{q}{m}\Bigr)_{\rm probe} \,\Bigl(\frac{q}{m}\Bigr)_{\rm background} \ga G_N$, and if the charged particles were a significant fraction of the cosmic contents, the gauge repulsion would affect the energetics of the universe and influence the expansion rate. Had the typical matter in the universe been the hydrogen atoms, with masses $m \sim {\rm GeV}$, the gauge repulsion would have significantly affected the cosmic balance for net monopole charges $q \sim m/M_{Pl} \sim 10^{-18}$, as noted in the theory of Lyttleton and Bondi \cite{bondi}. We can now recognize that this idea was really stemming from the mystery of nature which we have since identified as the gauge hierarchy problem: the tremendous disparity between the strengths of gravitational and electromagnetic forces. In that day, this at least motivated people to seek ways to improve the bounds on $q$ in terrestrial experiments, culminating in the 
results of King \cite{king}, who set the bounds on the electromagnetic 
fractional charge difference between an electron and a proton to $q \la 10^{-22}$, rendering it cosmologically irrelevant and laying the Lyttleton-Bondi theory to rest for a time. 

A simple way to elude the King bounds and resurrect long range gauge theory cosmological effects is to take the dark sector to be charged under some $U(1)'$. Indeed, as discussed in \cite{ack}, the dark sector $U(1)'$ couplings can be significantly stronger than the King bounds: $q^{\prime 2} \sim \alpha' \la 10^{-3}$ as opposed to $\alpha^{\rm eff}_{\rm King} \la 10^{-44}$. 
As already stated in the introduction, we now take our charged $U(1)'$ mode to be a TeV mass dark matter particle carrying integer units of dark charge $q' \sim g \sim 10^{-15}$, easily satisfying the dark sector bounds. If a significant fraction of dark matter is charged in this way, 
with similar fractions in extragalactic space and in galaxies and clusters, the gauge repulsion will affect the Hubble expansion as measured by the recession of faraway objects, because it will cancel the attraction between the charged masses. 

Indeed, consider Eq. (\ref{friedmann}) applied to the case where the universe is filled with a cosmological constant, neutral heavy dark matter and some  charged heavy dark matter (with `heavy' merely meaning, nonrelativistic particles {\it now}, with masses $\ga$ TeV).  
At some redshift $z = 1/R(t) - 1$, where $R(t)$ is the cosmological scale factor (playing the role of the distance between an observer and a receding beacon) we have $\rho_{\rm total} = \lambda + \rho_{\rm neutral} + \rho_{\rm charged}$, with $\rho_{\rm neutral}$ and
$\rho_{\rm charged}$ both scaling as inverse spatial volume: $\rho(z) = \rho(0) (1+z)^3$. Substituting this in the formula for $H$ given in Eq. (\ref{friedmann}), we find
\be
H^2(z) = \frac{8\pi G_N}{3} \, \lambda +\frac{8\pi G_N}{3} \Bigl\{ \rho_{\rm neutral}(0) 
+ \Bigl[1 - \frac{\Bigl(\frac{q}{m}\Bigr)_{\rm probe}(z) \,
\Bigl(\frac{q}{m}\Bigr)_{\rm background}(z)}{ G_N} \Bigr] \,\, \rho_{\rm charged}(0) \Bigr\} (1+z)^3 \, .
\label{friedmannz} \ee
This shows how the charged dark matter contributions get suppressed by gauge repulsion as a function of distance, or redshift. The explicit redshift dependence in $\Bigl(\frac{q}{m}\Bigr)_{\rm probe}(z)$ and 
$\Bigl(\frac{q}{m}\Bigr)_{\rm background}(z)$ is not of microscopic nature, but simply notes that an observer utilizes intrinsically {\it different} measures of cosmic expansion over the cosmic distance ladder, with different charged matter contents. This is the {\it key} point in what follows. For example, note that at very large redshifts one could extract the information about $H(z)$ from CMB or baryon distribution variation, which to leading order are not directly affected by $U(1)'$ repulsion\footnote{Of course, the neutral particles will be {\it indirectly} affected by the charges, by the fact that the structures in the universe, through which the neutral particles propagate, will be influenced by the extra forces. That does {\it not} affect to leading order the calibration of the luminosity-distance relationship, aside from the effect of inhomogeneities which can be dealt with in the familiar manner. It will however provide additional signatures of charges and forces, which should be considered separately.}. Since these probes are not charged under $U(1)'$, we take  
$\Bigl(\frac{q}{m}\Bigr)_{\rm probe}(z)=0$. In contrast, at low redshifts one would use the luminosity-redshift diagram of Type Ia SNe, which reside in galaxies that may contain charged dark matter, and so feel gauge repulsion. In this case, our probe {\it is} charged, so  $\Bigl(\frac{q}{m}\Bigr)_{\rm probe}(z)$ will be  non-zero. Even more prosaically, it may occur that as cosmic structure begins to form from adiabatic uncharged primordial seeds, the fraction of charged dark matter in the collapsed structures will initially grow, having started from zero, as long as the linearized approximation is valid and attraction is due to the primordial seed fields. The `absorption' of charge will be quickly saturated by nonlinear effects, when the repulsion from the matter already in the structure diminishes the net force acting on the charged dark matter, while the mass will continue to increase, and so at very small redshifts one would expect a {\it decrease} of $q/m$. Clearly,  this would also give rise to some non-trivial $z$ dependence in the background charge to mass ratio,  $ \Bigl(\frac{q}{m}\Bigr)_{\rm background}(z)$. The task of determining the details of the evolution of $q/m$ is difficult, but nevertheless very important, and we plan to address it in detail in future work.

If one normalizes $H_0$ by large $z$ observations and re-expresses (\ref{friedmannz}) in terms of fractional dark sector densities $\Omega_i = \frac{\rho_i}{\rho_{cr}}$, where $\rho_{cr}(z) = \frac{3H^2(z)}{8\pi G_N}$  is the critical energy density of the universe, one finds
\be
1=  \Omega_\lambda(z) - \frac{\Bigl(\frac{q}{m}\Bigr)_{\rm probe}(z) \, \Bigl(\frac{q}{m}\Bigr)_{\rm background}(z)}{G_N} \, \Omega_{\rm charged}(z)  + \Omega_{\rm DM}(z)  \, .
\label{friedmannznorm} \ee
Now suppose that using a set of fixed probes, which may experience extra forces, one does not account for the presence of dark charges and their repulsion, but merely identifies their contributions as `not dark matter', lumping them together with the cosmological constant into `dark energy' according to
\be
\Omega_{\rm DE}(z) = \Omega_\lambda(z) - \frac{\Bigl(\frac{q}{m}\Bigr)_{\rm probe}(z) \, \Bigl(\frac{q}{m}\Bigr)_{\rm background}(z)}{G_N} \, \Omega_{\rm charged}(z)  \, .
\label{denorm} \ee
Applying this to the low $z$ measurements shows that the gauge compensation of charged matter contributions effectively decreases the dark energy contribution $\Omega_\lambda$, and so to fit the data
at lower redshift  one may to increase the phenomenological value of $\lambda$ by {\it hand} at low $z$.
That would change the phenomenological value of $\Omega_\lambda$ at high and low redshift. This can contaminate the extraction of the effective equation of state of dark energy from the data, since it may simulate the effect of increasing dark energy with decreasing redshift. Indeed, using the phenomenological 
definition of total dark energy $\Omega_{\rm DE}$ in Eq. (\ref{denorm}), we can obtain the effective equation of state, $w_{\rm DE~ eff}$,  by using the definition of the equation of state of any given component $1+w_i=-\frac{\dot \rho_i}{3H\rho_i}=\frac{1}{3}(1+z)\partial_z \ln (\Omega_iH^2)$. So using Eq. (\ref{denorm}), we find that the effective dark energy equation of state is given by 
\be
w_{\rm DE~eff} = -1 - \frac{\Omega_{\rm charged}(z)}{\Omega_{\rm DE}(z)}   \frac{\Bigl(\frac{q}{m}\Bigr)_{\rm probe}(z) \, \Bigl(\frac{q}{m}\Bigr)_{\rm background}(z)}{ G_N}  \, .
\label{effw} \ee
where we have substituted $w_{\lambda}=-1$, and $w_{\rm charged}=0$.
In the derivation of this equation we have ignored $z$-derivatives of $\Bigl(\frac{q}{m}\Bigr)_{\rm probe}(z) \, \Bigl(\frac{q}{m}\Bigr)_{\rm background}(z)$ since that dependence does not really represent an intrinsic variation of the background, and merely reflects a change of the geometric probe at different redshift, as we have already explained.  It is now evident that the misinterpretation of gauge repulsion as a dark energy effect could mimic $w_{\rm DE~eff} < -1$. This possibility has recently attracted considerable theoretical thought
(see e.g. \cite{menace,ckt}), and spurred much exploration of data (e.g. \cite{wlow}).

We have -- for simplicity -- assumed the `real' dark energy to be a cosmological constant, as noted above. 
If the `real'  dark energy weren't a cosmological constant, but had  equation of state, $w_{\rm DE~real}$, Eq. (\ref{effw}) would read
\be
w_{\rm DE~eff} =  w_{\rm DE~real}\left(1+ \frac{\Omega_{\rm charged}(z)}{\Omega_{\rm DE}(z)}   \frac{\Bigl(\frac{q}{m}\Bigr)_{\rm probe}(z) \, \Bigl(\frac{q}{m}\Bigr)_{\rm background}(z)}{ G_N} \right) \, .
\label{effw-realw} 
\ee

Note that this immediately follows from Eq. (\ref{denorm}). Indeed, as the universe expands and the charged dark matter density redshifts away, the misplaced negative term in Eq. (\ref{denorm}) becomes smaller, and so the `net' dark energy $\Omega_{DE}$ appears to grow even faster. We stress again that the key for this
effect is clearly the use of different probes of cosmic expansion 
at different $z$, which have intrinsically different $\Bigl(\frac{q}{m}\Bigr)_{\rm probe}$, such that one may misinterpret the gauge repulsion as extra `anti-gravity'. 
Without this one would merely renormalize the  dark energy and dark matter contributions to the 
Friedmann equation relative to their mass fractions in the universe, as is clear from the equal 
scaling  with $z$ of both the neutral and charged dark matter contributions, for a fixed value of 
$\Bigl(\frac{q}{m}\Bigr)_{\rm probe} \, \Bigl(\frac{q}{m}\Bigr)_{\rm background}$. The point is that gauge repulsion contributions scale the same way as neutral dark matter ones, as is manifest in Eq. (\ref{friedmannz}). Without at least some change of the charge-to-mass ratio of the probe of the expansion and/or the background, the gauge repulsion alone could not affect the effective equation of state of dark sector\footnote{In their 1959 paper, Lyttleton and Bondi wanted gauge repulsion to drive the growth of cosmic scale factor. To get it, they were assuming a constant charged particle number density, resorting to the framework of the Steady State universe, where matter is continuously produced out of nothing. Without this assumption, the redshift of number density precludes the change of expansion rate, as we have seen here.}, or support a mirage of cosmic acceleration for long enough.

\section{From General Relativity to Faradayan cosmology}

The Faradayan cosmology description which we have used to write down Eq. (\ref{friedmann}) may appear  dubious. After all, the standard cosmological model is firmly grounded on General Relativity. The Newtonian approach however captures the essence of its physics correctly \cite{bondih,alvarez,jordan}. In the case of interest to us, this can readily be seen from an {\it exact} solution of Einstein's equations describing a system of charged massive particles in any FRW environment, where the charged particle masses are exactly compensated by their gauge repulsion. It is a generalization of the McVittie solution \cite{mcvittie} which includes any $U(1)$ charges \cite{vaidya},
and in the extremal limit where the charge is equal to the mass, it extends to a multicenter configuration thanks to a `no-force' condition for stationary particles with $q/m = \sqrt{G_N}$~\cite{kastra}. Taking the cosmological spatial sections to be intrinsically flat, the solution is given by \cite{vaidya,kastra}
\ba
&& ~~\,~~ ds^2 = - \frac{dt^2}{\Omega^2(t,\vec x)} + R^2(t) \Omega^2(t,\vec x) d\vec x^2 \, , 
~~~~~~~~~~~~~~~~~~~~~~ A = \frac{dt}{\Omega(t,\vec x)} \, ,  \nonumber \\
&& \Omega(t,\vec x) =1 + \sum_i \frac{\mu_i}{R(t) |\vec x - \vec x_i|} \, , ~~~~~~ 
\frac{\dot R}{R} = {H}(t) \, , ~~~~~~ {H}^2(t) = \frac{8\pi G_N}{3} \rho_{\rm neutral}(t) \, . ~~~~~~~~
\label{multicenter}
\ea
Here $\mu_i =  G_N m_i$ where $m_i$ is the mass of the source. 
The geometric asymptotia is controlled by the completely homogeneous neutral energy density 
$\rho_{\rm neutral}$, which may involve a cosmological constant, usual neutral dark matter, baryons etc.
The scale factor of the universe $R(t)$ obeys the standard Friedmann equation $H^2 = \frac{8\pi G_N}{3} \rho_{\rm neutral}$, with the charged dark matter contributions cancelled by the gauge repulsion terms thanks to $q/m = \sqrt{G_N}$ for all the charged masses in the system. Indeed, this no-force condition between any pair of charged masses guarantees that the charged dark matter only contributes to local distortions of the space-time geometry through $\Omega(t,\vec x)$, which is a harmonic function of the physical coordinates $\vec y= R(t) (\vec x - \vec x_i)$ . 

The solutions of the McVittie family in general FRW environs and their charged generalizations have been questioned on the basis of the singular structures in the geometry 
\cite{kras,nolan}. However, those issues only arise if one takes (\ref{multicenter}) to arbitrarily short distances. In the application we envision, the solutions (\ref{multicenter}) are naturally cut off by the finite size of the sources. As always, the cutoff scale is set by the inverse mass of the stable charged dark matter particle which sources (\ref{multicenter}), and which, by our discussion above, is bounded from below by $\ell_{\rm UV} \sim m^{-1} \ga \Lambda^{-1} \ga g^{-1} M_{Pl}^{-1}$ 
\cite{nima}. Thus the solution (\ref{multicenter}) should only be trusted down to distances 
$| \vec y| \ga g^{-1} M_{Pl}^{-1}$, where the potential function $\Omega$ is 
$\Omega \simeq 1 + {\cal O}(10^{-30})$, for the case of gravitationally coupled particles with $m \sim {\rm TeV}$. At shorter distances, one would have to use the full description of the matter source, which would regulate the solution just as the interior solution regulates the field of a planet at distances shorter than the planet's radius \cite{deser}. In any case, 
this will occur far far away from any strong gravity regime near a singularity, where the gravitational fields are practically still in the linear regime, and so whatever the issues involving McVittie and singularities, they need not concern us here.

\subsubsection*{\it Non-geodesic motion of charged particles}

Since different charged masses reside at fixed comoving coordinates $x_i$, their separation increases in time according to $\frac{d\vec y}{dt} = H \vec y$, which is precisely the Hubble law with $H$ set by only the neutral matter contributions. Indeed, if we consider a $t={\rm const}$ slice of (\ref{multicenter}) and evaluate the spatial distance between eg. the origin and a location of a charge at $\vec x = \vec x_i$, to leading order we find it to be given by the proper distance between the origin and the outer surface of the particle, $y = R |\vec x_i| - \ell_{\rm UV}  + \mu_i \ln(R |\vec x_i|/\ell_{\rm UV})$, where $\ell_{\rm UV}$ is the short-distance cutoff of the geometry near the charge. 
As time goes on, this distance increases according to 
$\frac{d y}{dt} = H\Bigl(y + {\cal O}(\ell_{\rm UV}, \mu_i, \mu_i \ln(y))\Bigr)$, with the time rate 
controlled by $H$ of Eq. (\ref{multicenter}), 
\be
H=\frac{\dot R}{R},  ~~~~~~ \qquad {H}^2 = \frac{8\pi G_N}{3} \rho_{\rm neutral} \, . 
\label{Hch}
\ee
Note, that these 
trajectories are {\it not} geodesics: they involve the $U(1)'$ Lorentz forces 
$\propto q_i \, u_\mu \, F'^{~\mu\nu}_j$ which encode the gauge repulsion 
of the charge $q_i$ by all the other charges in the system. This is why they 
can stay put at $\vec x_i = {\rm const}$, and move relative to each other 
with recession velocities set up only by the neutral mass density. In other words, 
the charged mass density contributions cancel exactly in (\ref{multicenter}), because both 
the charged sources and the probes have the same $q/m$, which satisfies 
$(q/m)^2 = G_N$, as anticipated by 
Eq. (\ref{friedmannz}). 

\subsubsection*{\it Geodesic motion of neutral particles}
On the other hand, neutral particles follow the geodesics of (\ref{multicenter}). The geodesic equations with the above notations read
\ba
 \Bigl(R^2 \Omega^2 \vec x~' \Bigr)' &=&  \frac{R^2 |\vec x~'|^2}{2} ~ \partial_{\vec x} ~ \Omega^2 -   
\frac{(t')^2}{2} ~ \partial_{\vec x} ~ \Omega^{-2} \, , \label{geodesic1} \\
 \Bigl(\frac{t'}{\Omega^2} \Bigr)' &=& \frac{(t')^2}{2} ~ \partial_t ~ \Omega^{-2} 
- \frac{|\vec x~'|^2}{2} \partial_t ~ \Bigl(R^2 \Omega^2 \Bigr) \, , \label{geodesic2} \\
 - \frac{(t')^2}{\Omega^2}+ R^2 \Omega^2| \vec x~'|^2  
  &=& - {\cal M}^2 \, , \label{geodesic3}
\ea
where the last equation follows from the affine reparameterization invariance of the geodesics, and 
${\cal M}$ is the integration constant that gives the affine parameter in the units of proper time. If we now follow the motion of a probe particle over sufficiently short times $T < H^{-1}$, such that $R \simeq {\rm const}$, then Eq. (\ref{geodesic2}) implies that  $t'/\Omega^2 \simeq {\rm const}$ over the course of this observation. Picking this ratio to be $t' /\Omega^2 \simeq {\cal M}$ so that a particle infinitely far from the massive sources can remain at rest in comoving coordinates ({\it i.e.} ensuring that we get 
$|\vec x~'| \to 0$ as $\Omega \to 1$), we obtain from Eq. (\ref{geodesic3})   
\be
R^2 \Omega^2 |\vec x~'|^2 \simeq {\cal M}^2 (\Omega^2 -1) \simeq 2 {\cal M}^2  V  \, . \label{pecvel}
\ee
to  leading order in the harmonic sum $ V = \sum_i \frac{\mu_i}{R |\vec x - \vec x_i|}$.
The ratio of  $R^2 \Omega^2 |\vec x~'|^2$ to $(t')^2/\Omega^2$ gives the comoving kinetic energy per unit mass of a neutral particle, or equivalently, the square of the comoving velocity of a neutral particle, 
\be
|\vec {\cal V}|^2 = \frac{R^2 \Omega^4|\vec x~'|^2}{(t'^2)} \simeq \Bigl(1- \frac{1}{\Omega^2} \Bigr) \simeq 2  V  \, . \label{pecvel2}
\ee
So the total velocity of a neutral particle is $\vec v_{\rm total} = \vec v_{\rm Hubble} + \vec {\cal V}$, where  $\vec v_{\rm Hubble}$ is the Hubble flow relative to our initial observer, and  $\vec {\cal V}$ is the total `peculiar' velocity of the probe. 

We now make some simplifying assumptions: (i) we ignore the asymmetries in the nearby mass distributions and treat the background as spherically symmetric about any observer's vantage point, and (ii) we view the probe's motion as being `virialized' with all the sources acting on it, such that its motion corresponds to merely marginal binding with the masses surrounding the observer. The latter approximation randomizes the  peculiar velocity of the probe, which is reasonable in order to extract the leading order bulk velocity.  Indeed, the `average projection' of the net velocity on the direction of the Hubble flow relative to our initial observer, which due to our virialization assumption means that the observer gathered the information from many probes in a patch of the sky with random direction of their peculiar velocity, to leading order yields $\langle \vec v_{\rm Hubble} \cdot \vec {\cal V} \rangle = 0$. Hence the net velocity squared of a particle at a distance
$R$ comparable to the horizon scale $H^{-1}_0$ (i.e. that corresponds to the comoving distance $|\Delta \vec x| \simeq 1$) from an observer is 
\be
|\vec v_{\rm total}|^2 =|\vec v_{\rm Hubble}|^2 + |\vec {\cal V}|^2 \simeq H^2R^2+2V \, .
\label{vtotal2}
\ee
Now imagine that the extra term in Eq. (\ref{vtotal2}) is interpreted by the observer  as a correction to the Hubble recession. It follows that  a neutral particle, with $q/m = 0$, will `see' an effective Hubble parameter $H_0$ given by
\be
H^2_0 \simeq H^2 +  \frac{2 V}{R^2} \, .
\label{neutralhubble}
\ee
Taking the continuum limit of the charged particle distribution and recalling that  $\mu_i = G_N m_i$, where $m_i$ is the mass of the individual charged dark matter particle, we see that $ \frac{2V}{R^2} = \Bigl(\sum_i \frac{2 \mu_i}{ R^3 |\Delta \vec x|}\Bigr)_{|\Delta \vec x |\simeq 1} \rightarrow \frac{8\pi G_N}{3} \rho_{\rm charged}$, where as noted before we evaluate the potential at the comoving scale corresponding to the horizon distance, $|\Delta \vec x| \simeq 1$. Substituting into (\ref{neutralhubble}) and using $H^2 = \frac{8\pi G_N}{3} \rho_{\rm neutral}$, as per Eq. (\ref{multicenter}), finally yields
\be
H^2_0 \simeq H^2 +  \frac{8\pi G_N}{3} \rho_{\rm charged} = \frac{8\pi G_N}{3} \rho_{\rm total} \, .
\label{neuthubble}
\ee
This is  precisely our starting Eq. (\ref{friedmann}) with $(\frac{q}{m})_{\rm probe} = 0$, as it should be for the case of a neutral probe. 

For both charged and uncharged probes, we have now seen that the linearized limit of the exact solutions precisely confirms the Newtonian cosmology argument, yielding the expected contribution from gauge repulsion. One  expects, on the basis of gauge symmetries of gravity and gauge theory, and the resulting energy and charge conservation, that the central features of this argument should remain valid beyond the linear limit (or more precisely, that there exists a gauge in which the theory effectively linearizes over the relevant scales, as for example in the derivation that led to (\ref{multicenter})). So indeed the neutral particles feel the attraction of the total mass in the universe, while the charged ones do not sense the attraction of the charged component due to its repulsive gauge force. 

\subsubsection*{\it Local cosmology from the Newtonian gauge potential}

So far we have obtained our `corrected' Friedmann equation 
(\ref{friedmannz}) using intuitive arguments based on the motion  of distant particles. Whereas the neutral particles follow geodesics, the charged particles do not, since they suffer an extra acceleration due to the gauge repulsion. In the previous two subsections, we explained how this gives rise to `probe dependence'  in the Friedmann equation (\ref{friedmannz}).  
To further allay any concerns that our `corrected' Friedmann equation 
(\ref{friedmannz})  has been derived `too heuristically', we will now reinforce the derivation of 
Eq. (\ref{neuthubble}) using an alternative approach, extracting it directly from the metric (\ref{multicenter}).  

Our aim now is to show that Eqs   (\ref{neutralhubble}) and (\ref{neuthubble}) follow from rewriting 
(\ref{multicenter}) in the `Newtonian gauge' as in, for example, \cite{gal}, and evaluating directly the Newtonian potentials at subhorizon distances. This is precisely consistent with our approach so far. So, let us begin with the full solution for the metric  (\ref{multicenter}) and expand about an observer at the origin, so that
\ba
ds^2 &\simeq& -\frac{dt^2}{\Omega_0^2}\left(1-\frac{2\vec x \cdot \vec k}{\Omega_0}+x^ix^j\left(\frac{3k_ik_j}{\Omega_0^2}-\frac{k_{ij}}{\Omega_0}\right)\right) \nonumber \\
&&\qquad\qquad + R^2(t) \Omega_0^2d\vec x^2\left(1+\frac{2\vec x \cdot \vec k}{\Omega_0} +x^ix^j\left(\frac{k_ik_j}{\Omega_0^2}+\frac{k_{ij}}{\Omega_0}\right)\right) +{\cal O}(x^3) \, ,  \label{pert}
\ea
where
\be
\Omega_0(t)=\Omega(t, \vec 0), \qquad k_i(t)=\frac{\partial \Omega}{\partial x^i}\Bigg|_{(t, \vec 0)} , \qquad k_{ij}(t)=\frac{\partial^2 \Omega}{\partial x^i\partial x^j}\Bigg|_{(t, \vec 0)} \, , 
\ee
Note that we do not need to worry about any of the gradients diverging since we have introduced a natural cut-off, $\ell_{\rm UV}$, around the core of each particle.
Clearly this metric may now be thought of as a local cosmological perturbation about the background Hubble flow set-up by the neutral matter content, $\rho_{\rm neutral}(t)$. 

In the absence of this perturbation, one can extract the background Hubble parameter by writing the FRW metric as a perturbation about the observer's local Lorentz frame on sub-horizon scales (see, for example, \cite{gal}). This is done by going to `Newtonian gauge', where the Hubble parameter is given in terms of the Laplacian of the Newtonian potential, evaluated at the observer's position, $H^2=\frac{2}{3} \vec \nabla^2 \psi|_{\rm obs}$. Clearly, a perturbation about the cosmological background complicates things, but only slightly. To proceed with the same method of evaluating the Hubble parameter as in the absence of local inhomogeneities, we will again rewrite our perturbed solution (\ref{pert}) as a sub-horizon perturbation about the observer's local  Lorentz frame, in Newtonian gauge. We begin with the coordinate transformation $(t, \vec x) \to (\tau, \vec y)$, where
\ba
t&=&T(\tau)-\frac{1}{2}H(T)\Omega_0(T)|\vec y|^2-\frac{1}{2}H(T)|\vec y|^2\frac{\vec y \cdot \vec k (T)}{R(T)\Omega_0(T)}+{\cal O}(y^4) \, ,  \\
\vec x&=& \frac{\vec y}{R(T)\Omega_0(T)}\left[1+\frac{1}{4}H^2(T)|\vec y|^2\right]-\frac{1}{R(T)^2\Omega_0(T)^3}\left[(\vec y \cdot \vec k (T))\vec y-\frac{1}{2}|\vec y|^2 \vec k(T)\right]\nonumber \\
&& +\frac{1}{R(T)^3\Omega_0(T)^5}\left[ \left((\vec y \cdot \vec k (T))^2-\frac{1}{4} |\vec k(T)|^2|\vec y|^2\right)\vec y-\frac{1}{2}(\vec y \cdot \vec k (T))|\vec y|^2\vec k(T)\right]+{\cal O}(y^4) \, , ~~~~~
\ea
and $T'(\tau)=\Omega_0(T)$. This immediately puts the metric (\ref{pert}) in the desired form,
\be
ds^2 \simeq -(1+2\phi) d\tau^2+(1-2\psi) d \vec y^2 \, ,
\ee
where the Newtonian potentials are given by
\ba
\phi&=&-\frac{1}{2}(H'(T)+H^2(T))|\vec y|^2-\frac{\vec y \cdot \vec k (T)}{R(T)\Omega_0(T)^2}
\nonumber\\&&\qquad +\frac{1}{2R(T)^2\Omega_0(T)^4}\left[5(\vec y \cdot \vec k (T))^2 -|\vec k(T)|^2|\vec y|^2\right]-\frac{y^i y^j k_{ij}(T)}{2R(T)^2\Omega_0(T)^3} +{\cal O}(y^3) \, ,  \\
\psi &=& \frac{1}{4}H^2(T)|\vec y|^2+\frac{1}{R(T)^2\Omega_0(T)^4}\left[(\vec y \cdot \vec k (T))^2 -\frac{1}{4}|\vec k(T)|^2|\vec y|^2\right] \nonumber\\
&&\qquad -\frac{y^i y^j k_{ij}(T)}{2R(T)^2\Omega_0(T)^3}+{\cal O}(y^3) \, .
\ea
These equations describe the gravitational field in the vicinity of the observer, as felt by neutral probes following geodesics of the metric (\ref{multicenter}). 
To extract the effective Hubble parameter measured by the observer recording the recession of these probes, we simply take the Laplacian of the Newtonian potential, $\psi$, as in \cite{gal}, and evaluate it at the observer's location. This yields
\be
H_0^2=\frac{2}{3} \vec \nabla_{y}^2 \psi|_{\vec y=\vec 0}=H^2(T)-\frac{2\vec \nabla^2_x \Omega|_{(T, \vec 0)}}{3R(T)^2\Omega_0(T)^3}+\frac{\Big |\vec \nabla_x \Omega|_{(T, \vec 0)}\Big|^2}{3R(T)^2\Omega_0(T)^4} \, .
\label{newthub}
\ee
As expected, the background Hubble parameter gets shifted due to the charged dark matter components. To recover Eq. (\ref{neuthubble}), we recall  that $\Omega=1+V$, where $V=\sum_i \frac{G_N m_i}{R |\vec x - \vec x_i|}$, and $m_i$ is the mass of the dark matter particle at $\vec x_i$. Taking the continuum limit, the potential $V$ becomes an integral $V \to \int d^3 x' \frac{G_N\rho(\vec x')}{R|\vec x-\vec x'|}$, where $\rho(\vec x')$ is the local comoving density of the charged dark matter. It follows immediately that
\be
\vec \nabla_x \Omega|_{(T, \vec 0)}=-\int d^3 x'\frac{G_N\rho(\vec x') \vec x'}{R(T) |\vec x'|^3}, \qquad  \vec \nabla^2_x \Omega|_{(T, \vec 0)}=-4\pi G_N \frac{\rho(\vec 0)}{R(T)} \, .
\label{omegas}
\ee
We now make use of our simplifying assumptions from before, and ignore asymmetries in the nearby mass distributions, treating them as spherically symmetric about the observer. This eliminates the first derivative terms, $\vec \nabla_x \Omega|_{(T, \vec 0)} \to 0$ (which appear non-linearly in Eq. (\ref{newthub}) anyway, and hence really correspond  to higher multipole corrections in the expansion, and 
are essentially negligible for our purposes), so that
\be
H_0^2=H^2(T)+\frac{8 \pi G_N}{3}\left(\frac{\rho(\vec 0)}{R(T)^3\Omega_0(T)^3}\right)= \left(H^2(\tau)+\frac{8 \pi G_N}{3}\rho_{\rm charged}(\tau)\right)(1 +{\cal O}(V)) \, ,
\ee
where $\rho_{\rm charged}(\tau)=\rho(\vec 0)/R(\tau)^3$ is the physical energy density of the charged dark matter, to leading order. As stated earlier, the ${\cal O}(V)$ corrections  are never larger than $10^{-30}$. Thus we recover Eq. (\ref{neuthubble}), providing a rigorous check of our modified Friedmann equation (\ref{friedmann}) for neutral probes.

\subsubsection*{\it Validity of the multi-centered black hole solution}

So far, we have treated the distribution of the charged dark matter particles as essentially static in comoving coordinates, making use of the multi-centered black hole solution (\ref{multicenter}). That is of course unrealistic: the dark matter particles won't just stay put at some fixed values of
$\vec x_i$, but will move around with peculiar velocities, as large as ${\cal O}(100) \, {\rm km/s}$. Due to the different boost properties of gauge and gravitational conserved currents (the latter being of course the conserved stress energy tensor), the no force condition between different charged particles will be violated,
leading to net forces between them that are proportional to their relative velocity. Thus the whole array 
(\ref{multicenter}) will really be dynamical, being unstable to 
nonzero velocity perturbations, and with individual constituents forever bouncing around in each other's fields \cite{giru,eardley}. However, this needn't concern us too much here, as we will now explain. 

First of all, the dark matter particles are packed together rather closely: since their number density in the universe is $n \sim \rho_{\rm cr}/m \sim 10^{-12} {\rm eV}^4/m$, their interstitial distance is $\ell_{\rm DM} \simeq1/n^{1/3}  \simeq (m/{\rm eV})^{1/3}  \, {\rm mm}$. If the dominant component of dark matter is a TeV-mass WIMP,  there is a dark matter particle in every cube of side ten meters or so in the extragalactic medium. The density of charged dark matter in  extragalactic space could be more dilute. However if the mass of charged dark matter is also in the WIMP range, which would follow from  the conjecture of 
\cite{nima} and our taking the $U(1)'$ strength to be gravitational, their separation will also be within an order of magnitude of WIMP's, i.e. $\sim {\cal O}(10)$ meters. On the other hand, if they are black holes, with masses  $m \ga M_{Pl} \sim 10^{18} {\rm GeV}$, their interstitial distance in the extragalactic medium will be of the order  $1000$ kilometers. In any case, in the units of the Hubble scale, they are very close together. Clearly, their separation would be yet smaller in condensed structures, such as galaxies, where the matter density may exceed the uniform extragalactic density by a billion times and more, being 
$\rho_{\rm galaxy} \ga 10^{-3} \, {\rm eV}^4$ or so. Thus, when considering the dark matter interactions, to leading order we can safely treat them as swarms of particles in flat space. 

In this case, the analysis of \cite{eardley} shows that they will affect each others trajectories, owing to the fact  that gravity has infinite range. However, they will  gravitationally coalesce only when the impact parameters are of the order of their horizons. If they are WIMP-scale stable particles, the dynamics of the theory which completes their description in the UV will become important at much longer scales, $\ell_{UV} \sim m^{-1}$ as per the conjecture of \cite{nima}. Even on the rare occasions where they form black holes with masses $M \gg m$, the weak gravity conjecture of \cite{nima} suggests that such black holes will decay back into the charged constituents, being merely metastable transients. Generically, however, such dark matter particles will simply zip around each other, mostly missing each other. In any case, at very large distances, to observers moving away due to Hubble recession, they will appear as nearly static swarms of mass and charge, whose conservation will justify using (\ref{multicenter}) as the appropriate 
leading order approximation\footnote{In the papers 
\cite{shiraishi1} the statistical mechanical description of the gas of charged black holes has been studied, with further interesting possibilities for cosmological applications found to occur if the gas condenses. Our analysis as exhibited in this work is appropriate for the dilute gas limit. Here, we will not address the short interstitial distance approximation where condensation may occur because of possible model dependence of the cutoff effects, that may come in because of the weak gravity conjecture of \cite{nima}, on the results of \cite{shiraishi1}. This remains an interesting question to be considered elsewhere.}. Indeed, as long as the distance to the observer exceeds $L \sim v_{\rm peculiar}/H$, such that the Hubble velocity is the dominant driver of the change in relative separation, it is perfectly reasonable to make use of the multi-centered black hole solution (\ref{multicenter}).

\section{Discussion}

We have clearly shown that the presence of dark matter charged under a dark $U(1)'$ can lead to a mismatch in the estimates of the dark matter density, depending on whether we use neutral or charged probes. Observers who fail to take into account the gauge repulsion between the galaxies which house type Ia SNe, and may contain charged
dark matter, will misinterpret the enhanced cosmic recession  as an otherwise mysterious `anti gravity' effect, inferring a smaller equation of state for dark energy than is actually the case. For example, if dark energy really were a cosmological constant, they  may erroneously deduce that the dark energy equation of state would lie beyond the phantom divide, $w<-1$. Although we have focussed on the role of probes used in observations, our corrected Friedmann equation (\ref{friedmannz}) includes the possibility that the background charge-to-mass ratio can also be redshift dependent. This can occur because  more charged dark matter will have been attracted to collapsed structures at later times. 

In the presence of net charge and associated gravitational strength long range forces 
there can be other effects. Let us comment on one very interesting effect that may also arise from inhomogeneities in our model: local variations in the density  of charged dark matter can lead to local variations in the Hubble flow. This could be relevant for explaining the recent claims of  \cite{dale}.    
Suppose that in some region of space, of size considerably smaller than the Hubble length, but big enough to accommodate many galaxies, there is an overdensity of charged particles $\delta \rho$. 
The presence of extra charged matter will lead to an extra repulsive force acting on the particles in that region, as compared to the forces acting on neutral particles, directed radially outward from the center of the overdensity. The velocity of the charged probes inside this region will therefore be larger than the mean recession velocity estimated by the Friedmann equation (\ref{friedmannz}). To get the scale of the effect, we can proceed as follows.  From Eq. 
(\ref{pecvel2}) we see that the charged probes will move a bit faster than the neutral ones, by an amount 
\be
\Delta v \simeq \sqrt{\vec {\cal V}^2} \simeq \sqrt{2 V} \, .
\label{delv}
\ee
Now compare this  to the mean recession velocity in a region of space of size $R$, 
given by $v \simeq HR$. Assuming that inside the overdensity the charged particles are distributed approximately uniformly, with density $\delta \rho_{\rm charged} = \xi \, \rho_{cr}$,  where $\rho_{cr}$ is the critical density, we find 
\be
\frac{\Delta v^2}{v^2} = \frac{\Delta H^2}{H^2} \simeq \frac{6 V}{8\pi G_N \rho_{cr} R^2} \simeq 
\frac{ \delta \rho_{\rm charged}}{\rho_{cr} }= \xi \, .
\label{delv1}
\ee
This means that the excess velocity will scale as the square root of the galactic charged mass density, in  units of the critical density. 

In the recent paper \cite{dale} it has been reported that there may be evidence for anomalous flows in an observed region of the universe of a typical scale $\sim {\cal O}(100) \, {\rm MPc}$, out to about $300$ MPc. Since $v^2 \simeq \frac{8\pi}{3} G_N \rho_{cr} R^2 \simeq H^2 R^2$, its numerical value at the scale of $R \simeq 100-300$ MPc, which is about $0.01$ to $0.03$ of the Hubble length, ranges between $v^2 \simeq 10^{-4}$ and $10^{-3}$. Thus, the relevant excess velocity is given by
\be
\Delta v^2 \simeq \Bigl (10^{-4} - 10^{-3} \Bigr) \cdot \xi \, .
\label{excess}
\ee
On the other hand, \cite{dale} report excess velocities of the order of $1000 \, {\rm km/sec}$, which, in units where $c=1$,  yields $0.003$. This gives a value for  $\Delta v^2$ of about $10^{-5}$. To reproduce this scale, the charge overdensity $\xi$ need only be in the range of 10 percent to a percent, diminishing as we move away from its center. This appears perfectly plausible, allowing for a local interpretation of the claims of \cite{dale}, if they are proven to be a real physical effect, and without any need of invoking superhorizon physics. It may therefore warrant a closer study and comparison with the data. 

Further, one may also expect that to subleading order, the CMB peaks could be affected because the charges will alter the formation of structure. The charges of the same sign will lead to a suppression os structure formation, since they will `screen' part of the collapsed structure's mass and impede its further growth. This can shift the CMB peaks some. To get an idea by how much, we note that if $\xi$  is the fraction of charged dark matter, the correction in the suppression of structure must go as $\xi^2$, just because the extra force requires two charges to operate, if we taking the initial density 
perturbation is adiabatic. In this case, the suppression will only occur as the additional charged dark matter particle are pushed away by the charges already inside the structure. This is also evident in our equations where the effects of charge go as $(q/m)^2$. Hence, small amounts of charge will not affect CMB a lot.

One should also worry about the possible effects of the extra $U(1)'$ on nucleosynthesis. If too much energy in the early universe is deposited in the massless $U(1)'$ mediators, it could compete with the photon energy during nucleosythesis, spoiling light element abundances. Indeed, we know that the extra light matter contributions to the expansion rate of the universe during nucleosythesis must be below photons by at least about two orders of magnitude. The question then is, how much $U(1)'$ will be produced early on? Clearly the answer is model-dependent; however, generically due to the weakness of the $U(1)'$ couplings, one expects the production of $U(1)'$ to be suppressed. Suppose, that the net charge in the universe is produced by the decay of an inflaton condensate that was charged under $U(1)'$. There will be {\it no} significant $U(1)'$ particle production during inflation itself, due to the conformal invariance of any $U(1)$ sector, well known from the example of Maxwell's theory in inflation. At the end of inflation, however, when the inflaton starts to oscillate around its minimum, there may be particle production due to the inflaton-$U(1)'$ coupling
$\sim g A_\mu J^\mu$, where $g$ is the $U(1)'$ coupling and $J^\mu$ is the inflaton current. If the spatial fluctuations of the inflaton produce a net $U(1)'$ dipole over a Hubble volume at the end of inflation, the total $U(1)'$ power will be of the order of $P \sim \omega^4 |p|^2 \sim H^4 \times q^2/H^2$, since the 
characteristic frequency is given by the inflaton mass, $\omega \sim m \sim H$, and the separation of charge in the dipole moment is maximized by the Hubble scale, $l \sim 1/H$, so that $p \sim q/H$. The energy density deposited in $U(1)'$ mediators will be $\rho_{U(1)'} \sim PT_H/V_H \sim q^2 H^4$. The net charge in the Hubble volume will be comparable to its net mass $M_H \sim \rho V_H \sim (G_N H)^{-1}$, 
and so $q\la 1$, implying that $\rho_{U(1)'} \sim H^4 \ll \rho_{reheating}$, because it is comparable to the gravitational particle production which is negligible when inflation occurs at, or below, the GUT scale.
Similarly, the repopulation of the universe by the $U(1)'$ mediators, that can occur by the collisional emissions of the $U(1)'$ quanta, will be suppressed relative to the photons by the square of the ratio of their couplings, $\sim g^2$, and it will be negliglble for gravitational couplings.

Thus, the most interesting question one can ask here is, just how much charge could there be? The answer should involve a detailed survey of various cosmological phenomena, and to have it we need to know the precise distribution of charges in the universe. Evidently, it is impossible to write down the formula for $(q/m)(z)$ without a detailed study of how structures form, and acquire charged dark matter. In the presence of charged dark matter, however, it must be non-zero for low red-shift observations like type Ia SNe, simply because when structures form by gravitational instability, some of the charged dark matter will fall in along with neutral matter. Thus, at least in the linear regime, one expects $q/m$ to be comparable to the total charged dark matter fraction at low redshift. Before this epoch, at large redshift, the charged particles will be distributed approximately uniformly in the universe. If during that epoch we use neutral probes, like CMB as observed by WMAP, which involve only the propagation of photons of zero charge-per-mass in the background geometry set up by total masses, to the leading order they will not directly discern the presence of the charged particles.  Clearly, at subleading order there will be effects, since the charges affect mass distributions which in turn affect the CMB anisotropies. Of course, to be able to estimate these effects, one should {\it i)} produce a precise parameterization of $(q_{eff}/m)(z)$ for a given class of cosmological probes, beyond the simple `step function' distribution sketched herein, and {\it ii)} account for possible systematic effects involving various uncertainties in present data. This task 
is beyond the scope of the present work, and we hope to return to it in the future. Here, our aim is to merely indicate that the search for anomalies in the velocity fields of the universe, cosmic structures and dark energy equation of state might reveal the presence of a new long range force, or at least place stronger bounds on them.

To this end, we can still get plausible numerical estimates of the scale of the effect. Imagine for simplicity that dark energy is a cosmological constant, 
with equation of state $w_{\rm DE~real}=-1$. Imagine also that the density of charged dark matter is, say, 
5\% of the real dark energy density, and take the $U(1)'$ strength to be exactly gravitational, assuming that the galaxies accreted charged dark matter in the same proportion to neutral matter, as in extragalactic space. From Eq. (\ref{effw}) one then gets that $w_{\rm DE~eff}=-1.053$. This is the dark energy equation of state inferred by an observer who does not know of the presence of dark charges, and correlates the low redshift supernova data as if all the large scale recession were controlled by gravity alone. As it stands, this might not appear as much. Yet note that it is near the limits of observationally testable deviations of dark energy equation of state from $w=-1$, and that it demonstrates how the data may suggest the crossing of the phantom divide without any exotic new physics such as ghosts or similar pathologies. Actually, the effect may be even larger if the dark 
$U(1)'$ is in fact stronger than gravity, such that $(q/m)^2>G_N$. This is consistent with the `gravity as the weakest force' conjecture of \cite{nima}, and may even be favored by it. In such a case, clearly, one would expect to have smaller $\Omega_{charged}$ in order to avoid disturbing too much the standard picture of structure formation, however this may be compensated by the increase in $G_N^{-1} (q/m)^2$. To get a more accurate idea of the possible range of $w_{DE~\rm{eff}}$, one would have to undertake a detailed study of bounds on charged dark matter from structure formation and other cosmological observations.

Note, that in a more accurate accounting for the effects of the charged dark matter population, one should also consider the influence of inhomogeneities in the cosmological backgrounds on the propagation of light. 
The inhomogeneities alone are known to affect the luminosity-distance relationship, perturbing the expression for angular diameter distance as a function of $z$. In a careful analysis this ought to be properly included \cite{dyer}. In the presence of charged dark matter, the inhomogeneities can arise from both neutral and charged particle density fluctuations, and should also be  carefully considered. We will ignore the detailed derivation of those effects here, keeping in mind that it can be done such that its influence is subtracted out of (\ref{friedmannznorm}) when considering the physical effects of dark charges. We hope to return to this issue more carefully in the future. 

In closing, let us summarize our findings. If there are dark matter particles which are charged under an unbroken $U(1)'$, of gravitational strength, they may have very interesting long range effects. Such states may arise in string theory, as the heavy BPS states. In the exact supersymmetric limit, their charges are equal to their mass, and, for states with masses close to the supersymmetry breaking scale, remain comparable even after the breaking\footnote{Corrections to the coupling will be negligible in the extreme weak coupling when $U(1)'$ is unbroken.}. We stress, that for the phenomena described here, the only required condition is that the charges, and the forces they source, are {\it comparable} to the particles' masses and their gravitational attraction, respectively; they do not need to be the same. Now, when their masses are of the order of string scale or greater, they are interpreted as extremal black holes. However, it has been conjectured \cite{nima} that in any UV complete framework gravity should be the weakest of forces, which implies a low cutoff in the gauge theory and the existence of stable states lighter than this cutoff. In any case, the long range fields of the charged dark matter can be described by the extremal black hole geometries below the cutoff, or equivalently outside of the sources. For particles with the same sign charges, their gauge repulsion will compensate gravitational attraction, weakening their net binding. If they partake in the matter population of the universe, measuring the masses with orbits of different probes, as is typically done in astronomy, may yield a mismatch between effective masses of cosmic objects. Specifically, we have seen that the flows of dark matter structures, such as galaxies, may yield smaller values if the gauge repulsion is not accounted for. In particular, the observation of type Ia SNe which reside in galaxies that contain charged dark particles, may misinterpret their faster cosmic recession due to weaker forces for an extra `antigravity', simulating effective dark energy equation of state smaller than the real one, and in some cases mimicking $w<-1$.
We have also noted that in the presence of charge overdensities, there may be  local variations of intergalactic forces, that can give a larger `Hubble Flow' in those regions of space. Such phenomena can be confused with a dynamical dark energy, or superhorizon effects. Thus it is important that such effects are taken into account when attempting to extract the properties of dark energy from cosmic motion and geometry, as they may be contaminated by the long range forces in the dark sector.

\vskip.5cm

{\bf \noindent Acknowledgements}

\smallskip

We would like to thank A. Albrecht, S. Chang, E Copeland, B. Freivogel, A. Green, D. Ko\v cevski, G. Niz, P. Saffin, L. Sorbo, D. Spergel, J. Terning and especially J.A. Tyson for valuable discussions and comments. We would also like to thank A. Flachi for providing us with a copy of \cite{vaidya}. AP thanks the UC Davis HEFTI program for hospitality during the inception of this research. NK thanks BIRS, Banff, Canada  for hospitality in the course of this work.
The work of NK is supported in part by the DOE Grant DE-FG03-91ER40674. 
The work of AP is  supported by a Royal Society University Research Fellowship.


\end{document}